\documentclass[12pt,preprint,apj]{emulateapj}

\usepackage{epstopdf}
\usepackage{color}
\usepackage{multirow}

\shorttitle{Fornax Ultra-faint Dwarf Galaxy}
\shortauthors{Lee et al. 2016}

\begin{document}

\title{
The Carnegie-Chicago Hubble Program: Discovery of the Most Distant Ultra-faint Dwarf Galaxy in the Local Universe}

\author{Myung Gyoon Lee\altaffilmark{1,}$^{\dagger}$, In Sung Jang\altaffilmark{1,2}, 
Rachael Beaton\altaffilmark{3}, 
Mark Seibert\altaffilmark{3},
%
Giuseppe Bono\altaffilmark{4},
and Barry Madore\altaffilmark{3}
}


%
%
%
\altaffiltext{$\dagger$}{mglee@astro.snu.ac.kr} 
\altaffiltext{1}{Department of Physics and Astronomy, Seoul National University, Korea}
\altaffiltext{2}{Leibniz-Institut  f{\"u}r Astrophysik Potsdam (AIP), An der
Sternwarte 16, D-14482, Potsdam, Germany}
\altaffiltext{3}{The Observatories of the Carnegie Institution of Washington, Pasadena, CA 91101, USA}
\altaffiltext{4}{Universit\`{a} di Roma  ”Tor Vergata”, Via della Ricerca Scientifica, 1 – 00133, Roma, Italy}


\begin{abstract}
Ultra-faint dwarf galaxies (UFDs) are the faintest 
known galaxies and due to their incredibly low surface brightness, it is difficult to find them beyond the Local Group.
We report a serendipitous discovery of an UFD, Fornax UFD1, 
in the outskirts of NGC 1316, a giant galaxy in the Fornax cluster. The new galaxy is located at a projected radius of 55 kpc   in the south-east of NGC 1316.
This UFD is found as a small group of 
resolved stars in the Hubble Space Telescope images of a halo field of NGC 1316, obtained as part of the Carnegie-Chicago Hubble Program.
Resolved stars in this galaxy are consistent with being mostly metal-poor red giant branch (RGB) stars. 
Applying the tip of the RGB method to the mean magnitude of the two brightest RGB stars, we estimate the distance to this galaxy, $19.0\pm1.3$ Mpc. 
Fornax UFD1 is probably a member of the Fornax cluster.
The color-magnitude diagram of these stars is matched 
by a 12 Gyr isochrone with low metallicity ([Fe/H] $\approx -2.4$).
Total magnitude and effective radius of Fornax UFD1 are $M_V \approx -7.6\pm0.2$    mag and $r_{\rm eff} = 146\pm9$ pc, which are similar to those of Virgo UFD1 that was discovered recently in the intracluster field of Virgo by Jang \& Lee (2014).
Fornax UFD1 is the most distant known UFD that is confirmed by resolved stars. 
This indicates that UFDs are ubiquitous and that more UFDs remain to be discovered in the Fornax cluster. 
\end{abstract}

\section{INTRODUCTION}

Ultra-faint dwarf galaxies (UFDs) are the faintest 
 known galaxies \citep{mcc12,bel13,mar16}.  Some of them are even fainter than typical globular clusters, 
 albeit they are much more extended 
 than the latter. Dwarf galaxies fainter than $M_V = -8.0$ mag and larger than  $r_{\rm eff}$  (effective radius) = 20 pc are often called as UFDs \citep{bec15}. They are optically feeble, but they are dominated by dark matter,  having very high values of mass to 
light ratios \citep{sim07,mcc12}. 
Thus nearby UFDs may be 
ideal laboratories to explore the particle properties of
dark matter (e.g., \citet{cal16}). 
Stellar populations in UFDs are mostly old metal-poor stars. \citet{bro14} found, from deep photometry of six UFDs 
around the Milky Way Galaxy, that formation of the stars in these galaxies
was finished by 11.6 Gyr ago ($z\approx 3$).
They suggested that this early quenching of star formation might have been due to global processes such as the reionization of the universe.
Thus, UFDs are considered to be fossils of the first galaxies \citep{bov11a,bov11b,jan14,ric16}.
Recently \citet{ric16} suggested, from high resolution simulations of the first galaxies, that some UFDs may be the remnants of a few distinct star clusters that were dissolved and expanded to the size of UFDs.


\begin{figure*}
\centering
\includegraphics[scale=1.0]{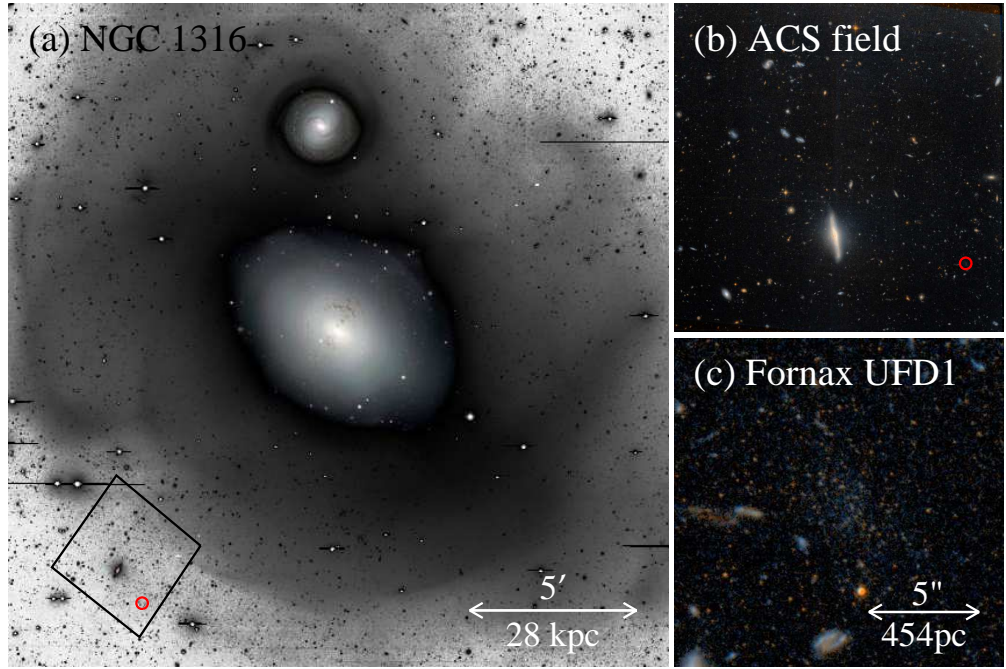} 
\caption{
(a) Location of the HST field (square) in the 
inverted color  image of  the inner region of NGC 1316 provided in the ESO archive, embedded in the gray scale map of 
$R$-band image given by \citet{ric12}. 
North is up, and east to the left. 
A small galaxy in the north of NGC 1316 is NGC 1317, a dwarf galaxy.
The position of the new UFD is denoted by a red circle.
(b) 
A color image of the HST field. 
(c) A $15\arcsec\times15\arcsec$ section of the HST field  centered on Fornax UFD1. 
Note that most resolved stars in the central region are red giants in this dwarf galaxy.
}
\label{fig_finder}
\end{figure*}

UFDs may be the most abundant type of galaxies in the universe, but they are revealing today only a tiny tip of a giant iceberg to our current observational facilities
\citep{mcc12,san12,bel14,lae14,jan14,dar16}.
They were first discovered, as a small  overdensity  
of resolved red giant stars, 
around the Milky Way Galaxy from the Sloan Digital Sky Survey (SDSS) data, 
and more of them were found later around M31 in the Local Group \citep{wil05a,wil05b,mcc12,san12,bel14,lae14,mar16}.
With recent advent of the Dark Energy Camera on CTIO 4m telescope, as many as 17 new UFDs 
have been discovered around the Milky Way Galaxy and the Large Magellanic Cloud (LMC) in 2015 \citep{bec15,kop15,kim15a,drl15,dar16}. It is expected that more UFDs will be discovered in the Local Group soon from current and future wide field surveys (e.g., LSST, PANStarrs, etc). UFDs in the Local Group can play an important role to help resolving the well-known missing satellite dwarf problem \citep{moo99} and understanding the nature of the first galaxies.
However, it is very difficult to find UFDs outside the Local Group with the current observational facilities. Therefore little is known about the UFDs outside the Local Group.

Surprisingly,  \citet{jan14} discovered one UFD, Virgo UFD1, in the intracluster field of the Virgo cluster, using deep images of the Hubble Space Telescope (HST) available in the MAST archive. 
Virgo UFD1 was found as a resolved stellar overdensity. 
Virgo UFD1 is the first UFD found beyond the Local Group. 
 Total magnitude and effective radius of Virgo UFD1 are $M_V \approx -6.5\pm0.2$ mag and $r_{\rm eff} = 81\pm7$ pc, and its central surface brightness is as low as $\mu_{V,0} = 26.37\pm0.05$ mag arcsec$^{-2}$.
Metallicity of the stars in this galaxy is estimated  from their colors, to be, on average, very low, [Fe/H] $=-2.4\pm0.4$.
Virgo UFD1 is located at the distance of $16.4\pm0.4$ Mpc, being the most distant UFD at the time of discovery.
Later \citet{crn16} reported detection of one UFD
in the remote halo of NGC 5128 (Cen A), a massive elliptical galaxy at the distance of 3.8 Mpc. They used the images obtained as part of the Panoramic Imaging Survey of Centaurus and Sculptor (PISCeS) at the Magellan 6.5 m telescope.
  

In this Letter 
we report a serendipitous discovery of an UFD, Fornax UFD1, 
breaking the distance record of Virgo UFD1. 
It is located in the outskirts of NGC 1316, a giant galaxy in the Fornax cluster.
Fornax UFD1 was detected as a small group of resolved stars in the HST/ACS images for a halo field of NGC 1316, obtained as part of the Carnegie-Chicago Hubble Program (CCHP) \citep{bea16}.

\section{DATA}

The CCHP is an ongoing program to estimate the value of the Hubble constant ($H_0$) with high precision using the population II distance indicators,
using RR Lyrae and 
the tip of the red giant branch (TRGB) \citep{bea16}.
 As part of the CCHP, deep F606W and F814W images of a halo field in the outskirt of NGC 1316 using HST/ACS were obtained. Total exposure times  for F606W and F814W are 14,676s and 24,396s, respectively.
These data were used for estimation of the distance to NGC 1316 using the  TRGB \citep{lee93}, and
detailed analysis of the data is given in \citet{jan17}. We use photometry of the point sources derived from these images by \citet{jan17b}. 
The magnitudes of the point sources in the images were derived using DAOPHOT \citep{ste94}.
In this study we adopted a distance to NGC 1316 based on the TRGB, given in \citet{jan17}: $(m-M)_0 = 31.45\pm0.05{\rm (ran)}\pm0.06{\rm (sys)}$ and $d=19.5\pm0.5$ Mpc. 
Foreground extinction and reddening values toward NGC 1316 are small, $A_V= 0.057$ and $(E(V-I)=0.026$ \citep{sch11}.

{\color{red}\bf Figure \ref{fig_finder}(a)}  displays the location of the HST field in the $B, V, R$ and $H\alpha$ color image of the inner region of NGC 1316 provided by the ESO archive, embedded in the gray-scale map of the deep $R$-band image given by \citet{ric12}. 
A smaller spiral galaxy in the north of  NGC 1316 is NGC 1317 (SAB(r)a).
This image shows various substructures including shells and ripples 
 around NGC 1316, which are probably merger remnants \citep{sch80,ric12}.
The HST field is located at $9\farcm5$ south-east of the NGC 1316 center, where diffuse stellar light of NGC 1316 is barely visible.  It is expected that this HST field would be dominated by old halo stars in NGC 1316.  
A color 
image  of the HST field is also displayed in  {\color{red}\bf Figure \ref{fig_finder}(b)}. 

\begin{figure*}
\centering
\includegraphics[scale=0.9]{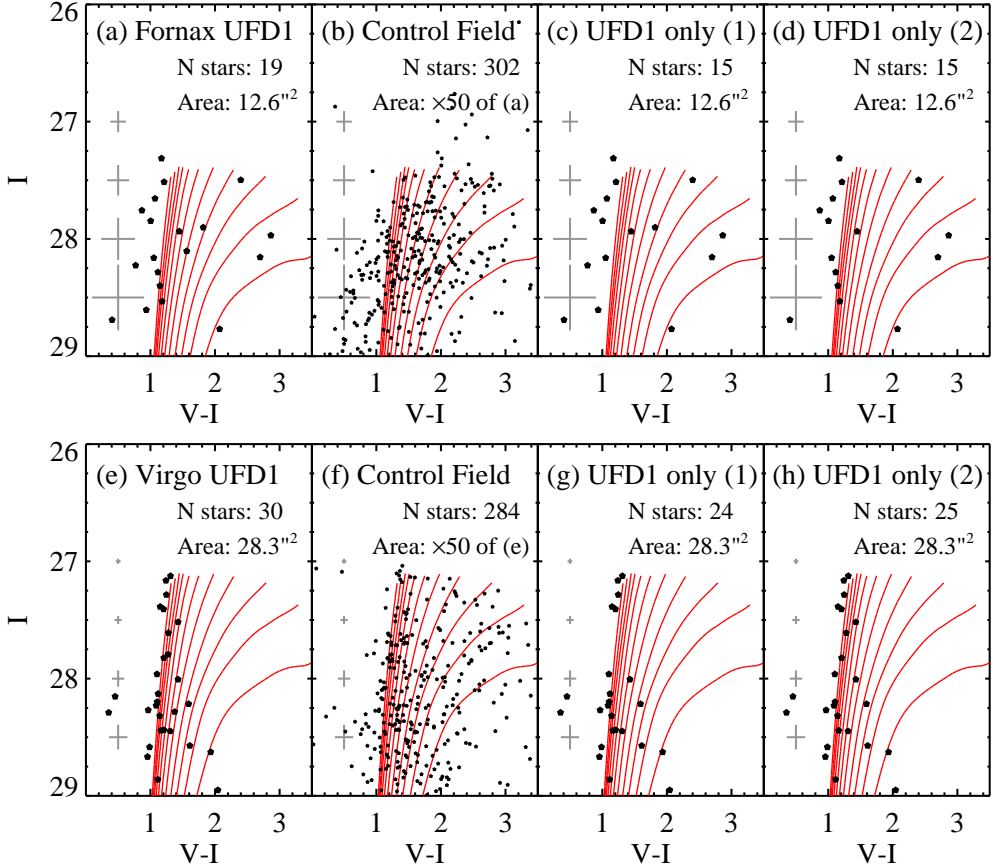} 
\caption{(Upper panels) $I - (V-I)$ CMDs of the resolved stars (a) at $r\leq2''$ of the Fornax UFD1 and (b) in  the control field at $5\farcs0 < r \leq 15\farcs0$. The area of the control field is 50 
times bigger than the UFD1 field. 
Mean values of the DAOPHOT errors are denoted by the error bars.
(c) and (d) The net CMDs of the UFD1 field obtained after statistical subtraction of the field contribution in two different realizations.
Curved lines denote the 12 Gyr stellar isochrones for  [Fe/H] = $-2.4$ to $-0.4$ in steps of 0.2 provided by the Dartmouth group \citep{dot08},  shifted in magnitude and color according to 
$(m-M)_0=31.37$ 
and $E(V-I)=0.026$.
(Lower panels) Same as the Upper panels but for Virgo UFD1.
}
\label{fig_cmd}
\end{figure*}

\section{RESULTS}

\subsection{Discovery of a New UFD}
First, we searched for any dwarf galaxies with low surface brightness  in the HST images of NGC 1316, which would appear to be similar to the morphology of Virgo UFD1 (see Fig. 1 in \citet{jan14}). Since the Fornax cluster is only slightly more distant than the Virgo cluster, it is expected that some of the brightest stars in Fornax UFDs, if any, would be resolved in the deep HST images.

Through visual inspection of the images, we noticed the presence of a small group of faint resolved point sources embedded in diffuse stellar light in the southern part of the HST field (at the position of the small red circle in {\color{red}\bf Figure \ref{fig_finder}(b)}). 
A zoom-in image of the field  centered on the new system 
in {\color{red}\bf Figure \ref{fig_finder}(c)}  shows clearly a faint fuzzy system embedding a small number of resolved stars, the diameter of which is 
about  $4''$.
This new system looks much fainter and smaller than any other typical dwarf galaxies in the Fornax cluster \citep{fer89,mie07,mun15}.
This system turns out to be, indeed,  a new UFD, named as Fornax UFD1, as shown below.
Fornax UFD1 is located $10\farcm1$ ($\sim$55 kpc) south-east of the NGC 1316 center in the sky.

\subsection{Color-magnitude Diagrams of the Resolved Stars in Fornax UFD1}

{\color{red}\bf Figure \ref{fig_cmd}(a)} displays the color-magnitude diagrams (CMDs) of the resolved point sources in a circular region 
 with $r\leq2''$ where $r$ is the projected radial distance from the center of Fornax UFD1. 
We plotted also the CMD of the control field in the annular region at $5''<r\leq15''$,
  in {\color{red}\bf Figure \ref{fig_cmd}(b)}.
The area of this control field is 
50 times bigger than that of the Fornax UFD1 field.
The CMD of the point sources in the control field shows a broad red giant branch (RGB), and these point sources are mostly RGB  stars in the halo of NGC 1316 \citep{jan17}.
We used statistical subtraction to remove 
the contribution of the  foreground and background sources from the CMD in (a), using the CMD of the control field.
 For this process we used bins with 0.3 mag in both magnitude and color, which is similar to the mean errors of the RGB stars.
{\color{red}\bf Figure \ref{fig_cmd}(c) and (d)} show two realizations of the control field subtraction. Both CMDs look very similar in general.
In the lower panels of {\color{red}\bf Figure \ref{fig_cmd}}, we plotted similar CMDs but for Virgo UFD1 \citep{jan14} for comparison.

\begin{figure*}
\centering
\includegraphics[scale=0.9]{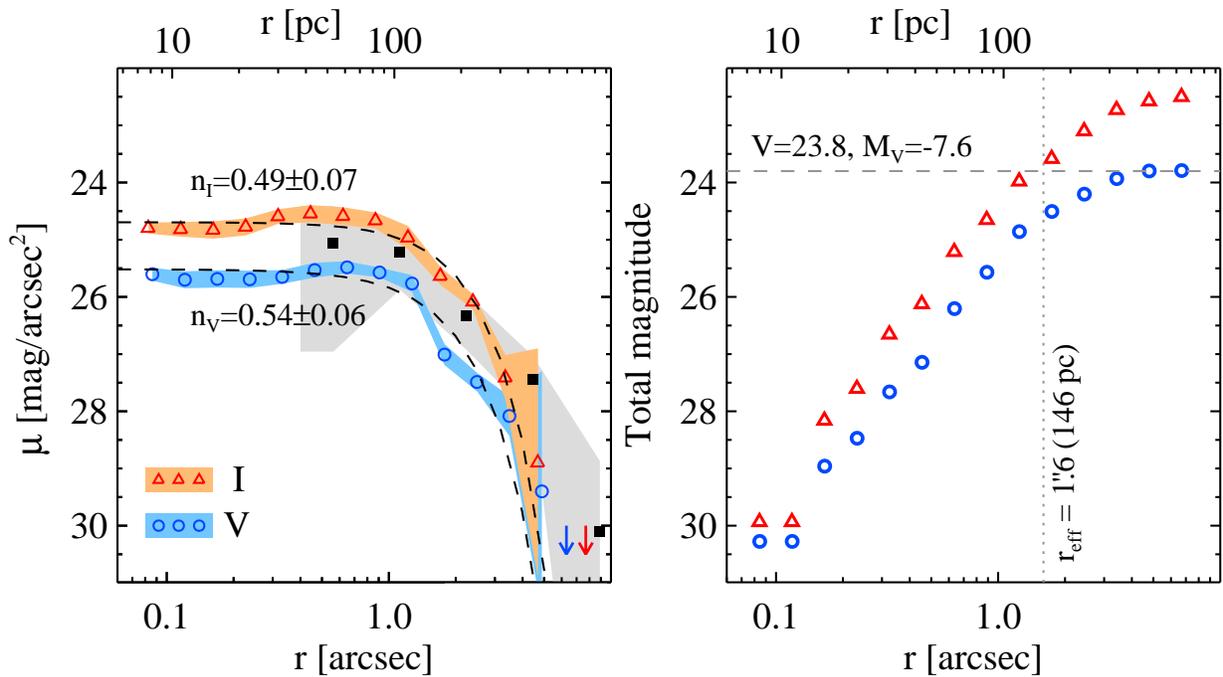} 
\caption{ 
The radial profiles of the $V$ (open circles) and $I$-band (open triangles)  surface brightness (a) and cumulative integrated magnitudes (b) of Fornax UFD1.
The arrows denote the location of the control region.
The bands represent the errors.
Filled squares with gray bands denote the radial number density profile of the resolved red giant stars ($27<I<28.5$ mag  and $0.0<(V-I)<2.0$).
Dashed lines in (a) represent a S\'ersic law fit for each band.
The $V$-band effective radius of Fornax UFD1 is marked by a vertical line in (b).
}
\label{fig_surf}
\end{figure*}

There are 12 resolved stars brighter than $I=28.5$ mag in the net CMDs ({\color{red}\bf Figure \ref{fig_cmd}(c) and (d)}).
The number of field contaminants with $I<28.5$ mag in the area covered by the UFD field is estimated to be $4.8\pm0.3$, 
while there are 15 stars in the area of UFD. Thus the net number of the member stars in the area of the UFD is estimated to be $10\pm4$, which is consistent with the number derived after CMD  decontamination. 
Eight of the  12 stars have a narrow range of colors with $0.7<(V-I)<1.4$, showing a vertical sequence, while three of them show much redder colors than the others. Therefore, these eight stars are considered to show the RGB of Fornax UFD1, while three redder stars are probably foreground or  background sources.

 Because of the small sample of the RGB stars in the CMD, it is not easy to derive an accurate distance to Fornax UFD1. We estimate approximately the distance to this UFD, assuming that the mean magnitude of the two brightest blue RGB stars in the CMD, 
corresponds to the magnitude of the TRGB.
The mean magnitude and color of these two stars are 
$I= 27.38 \pm0.14$ mag 
and $(V-I)=1.19\pm0.03$. 
 This magnitude value is 0.24 mag fainter than the value of Virgo UFD1, $I=27.14\pm0.04$ mag \citep{jan14}
 (see the lower panels in Figure \ref{fig_cmd}). 
Applying the TRGB calibration in \citet{jan17b} to  this TRGB magnitude, we derive a value for the distance to Fornax UFD1, $d=19.0\pm1.3$ Mpc 
($(m-M)_0 = 31.37\pm0.15$). 
This value 
is consistent with the distance to NGC 1316. 
This shows that Fornax UFD1 is likely to belong to Fornax. 

In {\color{red}\bf Figure \ref{fig_cmd}} we overlayed 12 Gyr isochrones for [$\alpha$/Fe]=0.2  \citep{fab15} and [Fe/H] $=-2.4$ to --0.4 in the Dartmouth model \citep{dot08}, which were shifted
according to the distance and foreground reddening for Fornax UFD1.
Although the number of stars is small, the RGB of UFD1 is, on average, bluer and narrower compared with the RGB of NGC 1316 in the control field. 
This shows that the mean metallicity of this new UFD is much lower than that of the NGC 1316 halo.
The CMD of these RGB stars is matched roughly by the RGB part of the isochrones for low metallicity.
It is noted that the RGB of Fornax UFD1 is even bluer than the RGB of the lowest metallicity with [Fe/H] $\approx -2.4$. 
We estimate roughly the mean metallicity  of the RGB stars to be [Fe/H] $\approx -2.4\pm0.4$. 
%


\subsection{Basic Parameters of Fornax UFD1}

We 
 constructed radial profiles of surface brightness and  cumulative integrated magnitude of Fornax UFD1 using IRAF/ELLIPSE, 
  as plotted in {\color{red}\bf Figure \ref{fig_surf}(a) and (b)}.
We masked out cosmic rays and background galaxies in the original image, and ran the ELLIPSE task adopting a smoothing option of $3 \times 3$ pixel binning.
The radial profiles of surface brightness in {\color{red}\bf Figure \ref{fig_surf}} show a flat core in the inner region at $r<1''$ and decline rapidly in the outer region, showing that this UFD is a dynamically relaxed system like typical old star clusters.
The radial profiles of surface brightness for $r<4''$ are fitted well by a Sersic law \citep{ser63} with an index $n=0.54\pm0.06$ for $V$-band and $n=0.49\pm0.07$ for $I$-band.
The effective radius measured at the $V$-band of this galaxy is 
$r_{\rm eff} = 1\farcs6\pm0\farcs1$ ($146\pm9$  pc). This value is much larger than that of the typical  globular clusters, $r_{\rm eff} \approx 3$ pc.
Its central surface brightness for $V$ and $I$-bands corrected for foreground extinction are derived to be $\mu_{V,0} = 25.46\pm0.06$ mag arcsec$^{-2}$ and
$\mu_{I,0} = 24.66\pm0.05$ mag arcsec$^{-2}$, respectively.
We derived the radial number density profile of the RGB stars with $I<28.5$ mag and $0<(V-I)<2.0$, choosing an annular region at $r \approx 20''$ to subtract the field contribution.
This profile at $r<5''$ is similar to the surface brightness profiles, as shown in {\color{red}\bf Figure \ref{fig_surf}}.

\begin{figure*}
\centering
\includegraphics[scale=0.9]{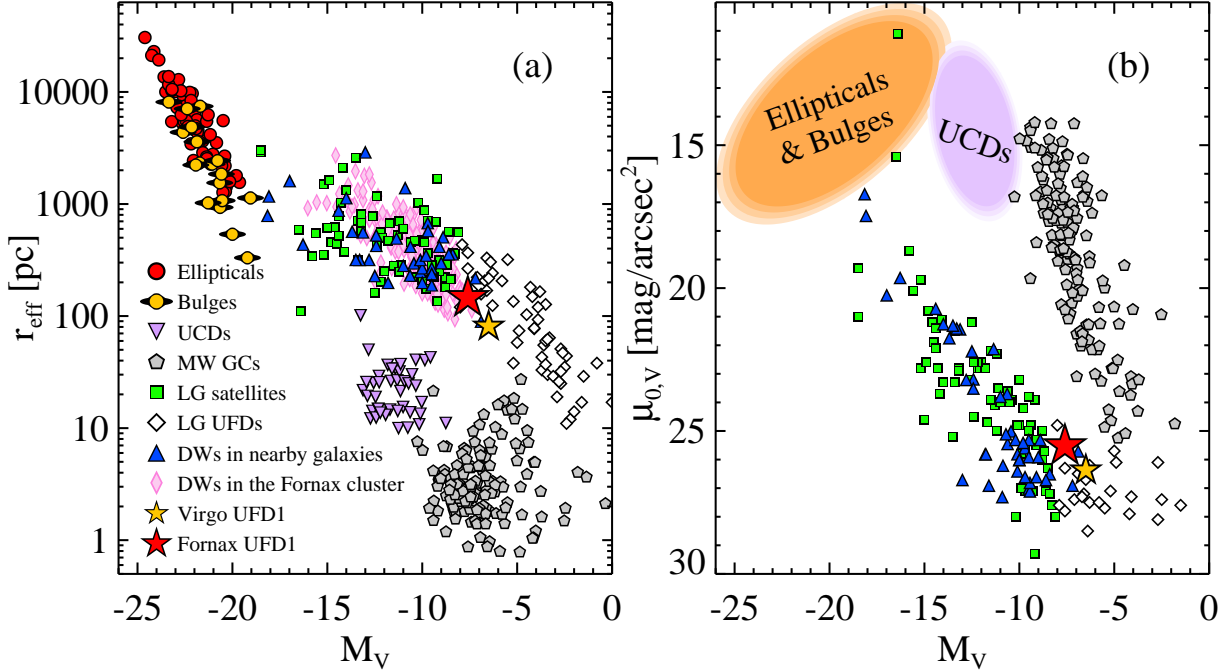} 
\caption{(a) The  effective radius versus  absolute $V$ total magnitude of the Fornax UFD1 (large red starlet symbol) in comparison with those for other stellar systems. We also plotted the Virgo UFD1 by the small yellow starlet symbol.
(b) The  $V$-band central surface brightness versus  absolute $V$ total magnitude of the Fornax UFD1.
Circles and lenticular symbols are for the giant ellipticals and bulges in spiral galaxies, 
downward triangles for the ultra compact dwarfs (UCDs), 
pentagons for the Milky Way globular clusters, 
squares and empty lozenges for the Local Group satellites and UFDs, 
and upward triangles for the dwarfs in M81 and M106 and the low surface brightness galaxies in M101, 
and pink diamonds for the NGFS dwarfs in \citet{mun15}. 
%
}
\label{fig_comp}
\end{figure*}

The integrated color  of Fornax UFD1 for the aperture of $r_{\rm eff}$ is estimated to be $(V-I) \approx 0.98\pm0.05$ ($(V-I)_0 \approx 0.95\pm0.05$).
The integrated color of Fornax UFD1 is consistent with those of  early-type dwarf galaxies in Fornax \citep{mie07}, and that of Virgo UFD1 \citep{jan14}.

The radial profiles of $V$ and $I$-band integrated magnitude
of this galaxy become approximately constant at $r = 5\arcsec$. The integrated magnitude for $r = 5\arcsec$ is estimated to be $V = 23.8\pm0.2$ mag 
and  $I = 22.5\pm0.2$ mag. 
These values can be considered to be the 
total magnitudes of Fornax UFD1.
Corresponding absolute magnitudes of this galaxy are derived to be $M_V = -7.6\pm0.2$ mag 
   and   $M_I =-8.9\pm0.2$ mag. 
   These results show that this new system is indeed a genuine UFD.
We list basic parameters of Fornax UFD1 in {\color{red}\bf Table 1}.
 
\section{Discussion} 

\subsection{Comparison with Previous Surveys of Fornax Dwarf Galaxies}

\citet{fer89} published the Fornax Cluster Catalog (FCC) based on the wide field photographic survey of Fornax galaxies. 
Following this, several studies used CCD images  to find new fainter galaxies in Fornax and investigate their properties \citep{kam00, hil03, mie07,mun15}.
\citet{kam00} 
imaged a long strip region from NGC 1399  to NGC 1291 (an area of 13.8 deg$^2$) in Fornax, finding a large number of low surface brightness dwarf galaxies with $M_B<-12.0$ mag. 
 These are unresolved faint galaxies.

Later deeper surveys of the central region in Fornax found a number of fainter dwarf galaxies. The detection limits of the dwarf spheroidal galaxies (dSphs) in the survey of \citet{hil03} and \citet{mie07} are
$V\approx 23$ mag ($M_V \approx -8.5$ mag), and $\mu_V (0) \approx 27$ mag arcsec$^{-2}$.
\citet{mie07} determined distances to early-type dwarf galaxies in Fornax using the surface brightness fluctuation (SBF) method, and confirmed that 30 of the dEs with $-16.6<M_V<-10.1$ mag are the  members of Fornax. 
%
Fornax UFD1 is about one magnitude fainter that the detection limits of \citet{hil03} and \citet{mie07}.

Recently \citet{mun15} covered the central region of Fornax with Dark Energy Camera on CTIO 4m as part of the program for the Next Generation Fornax Survey (NGFS), discovering 158 new dwarf galaxies with $M_i <-8.0$ mag. The effective radii and effective surface brightness of these galaxies are
$r_{\rm eff}=0.1$ to 2.8 kpc  and
$\mu_i = 22.0$ to 28.0 mag arcsec$^{-2}$.
Thus the parameters of Fornax UFD1 are similar to the faint limit in their sample.
However, none of the stars in Fornax UFD1-like dwarf galaxies can be resolved at the pixel scale of the Dark Energy Camera images. 


\begin{deluxetable*}{lcc}
\tabletypesize{\scriptsize}
\setlength{\tabcolsep}{0.05in}
\tablecaption{Basic Parameters of Fornax UFD1}
\tablewidth{0pt}
\tablehead{ \colhead{Parameter} & \colhead{Value} & \colhead{References}}
\startdata
R.A.(2000) 	& $3^h23^m11.^s374$		& 1		\\
Dec(2000) 	& $-37\arcdeg20\arcmin46\farcs53$ 	& 1 \\
Type		& UFD & 1								\\
Distance modulus, $(m-M)_0$ & $31.37\pm0.15$ & 1 \\ 
Distance, $d$ [Mpc] & $19.0\pm1.3$ & 1 \\
Image scale 				& 
91.1 pc arcsec$^{-1}$ & 1 \\
Total magnitude	& $V^T= 23.8\pm0.2$ 
& 1 \\
Color at effective radius	& $(V-I)_0 =0.95\pm0.05$ & 1 \\
Foreground reddening & $A_V=0.057$, $E(V-I)=0.026$ & 2 \\
Absolute magnitude, $M_V$	& $-7.6 \pm0.2$ & 1 \\ 
Metallicity & [Fe/H]$\approx -2.4$  & 1\\
Sersic Index (n), $V$ 		&
$0.54\pm0.06$ & 1  \\ 
Sersic Index (n), $I$ 		&
$0.49\pm0.07$ & 1  \\ 
Effective radius ($r_{\rm eff}$), $V$ 	& $1\farcs6\pm0\farcs1$ ($146\pm9$ pc) 
& 1    \\
Central surface brightness, $V$	& 
$\mu_{V,0}= 25.46\pm0.06$ mag arcsec$^{-1}$   & 1 \\
Central surface brightness, $I$	&  
$\mu_{I,0}=24.66\pm0.05$ mag arcsec$^{-1}$   & 1 \\
\enddata
\tablerefs{(1) This study; 
(2) \citet{sch11}.}
\label{tab_param}
\end{deluxetable*}

\subsection{Comparison with Other Galaxies}
We compare scaling relations of Fornax UFD1 with those of other early-type stellar systems in {\color{red}\bf Figure \ref{fig_comp}(a) and (b)}
 (see Figure 4 for details of the kinds of other early-type stellar systems). We use the data for effective radii, total magnitudes, and central surface brightness of other stellar systems compiled in \citet{jan14} (see the references therein). 
We updated the data, adding data for new UFD galaxies: the new UFDs in the Local Group from the Dark Energy Survey (DES)  \citep{bec15,kop15,drl15,kim15a,kim15b,lae15}, 
a new UFD in the Centaurus A group \citep{crn16} and Virgo UFD1 \citep{jan14}. 
We also added the data for the new Fornax dwarf galaxies \citep{mun15}.

In the figure it is found that Fornax UFD1 is located in the region of the Local Group UFDs and that the 
{\bf properties} of Fornax UFD1 are very close to those of Virgo UFD1 \citep{jan14}. 
These results show that Fornax UFD1 is indeed a genuine UFD.
The central surface brightness of Fornax UFD1 as well as Virgo UFD1 corresponds to the upper range for the Local Group UFDs. 
This implies that UFDs are ubiquitous and that there may be many more UFDs with lower surface brightness and fainter magnitudes in Fornax and Virgo.

\section{Summary}

Taking advantage of high-resolution deep images of a halo field in NGC 1316 obtained as part of the CCHP, we discovered a new UFD in Fornax, Fornax UFD1.
The CMD of the resolved stars in this UFD
shows that it is indeed a genuine UFD with low metallicity ([Fe/H] $\approx -2.4$)  and that it is likely to belong to the Fornax cluster.
The size, luminosity, and surface brightness of Fornax UFD1 are similar to those of Virgo UFD1. Fornax UFD1 is the most distant one among the known UFDs that have been confirmed by resolved stars. Fornax UFD1 and Virgo UFD1 are, respectively, only the first known UFD in each of Fornax and Virgo, and it is expected that
many more will be discovered in the future.
We just opened an window to the realm of the UFDs 
 beyond the Local Group.

\bigskip
The authors are grateful to anonymous referee for useful comments.
Tom Richtler is thanked for providing deep CTIO 4m images of NGC 1316. 
MGL and ISJ are supported by the National Research Foundation of Korea (NRF) grant
funded by the Korean Government (MSIP) (No. 2012R1A4A1028713).

Facilities: HST (ACS, WFC3/IR).

\clearpage

\clearpage


\begin{thebibliography}{}







\bibitem[Beaton et al.(2016)]{bea16} Beaton, R.~L., Freedman, W.~L., Madore, B.~F., et al.\ 2016, \apj, 832, 210 

\bibitem[Bechtol et al.(2015)]{bec15}Bechtol, K., Drlica-Wagner, A., Balbinot, E., et al. 2015, \apj, 807, 50





\bibitem[Belokurov(2013)]{bel13} Belokurov, V.\ 2013, NewAR, 57, 100  

\bibitem[Belokurov et al.(2014)]{bel14} Belokurov, V., Irwin, 
M.~J., Koposov, S.~E., et al.\ 2014, \mnras, 441, 2124 



\bibitem[Bovill 
\& Ricotti(2011a)]{bov11a} Bovill, M.~S., \& Ricotti, M.\ 2011a, \apj, 741, 17  

\bibitem[Bovill 
\& Ricotti(2011b)]{bov11b} Bovill, M.~S., \& Ricotti, M.\ 2011b, \apj, 741, 18 




\bibitem[Brown et al.(2014)]{bro14}Brown, T. M., Tumlinson, J., Geha, M., et al.\ 2014, \apj, 796, 91



\bibitem[Calabrese \& Spergel(2016)]{cal16} Calabrese, E., \& Spergel, D.~N.\ 2016, \mnras, 460, 4397 



\bibitem[Crnojevi{\'c} et al.(2016)]{crn16} Crnojevi{\'c}, D., Sand, D.~J., Spekkens, K., et al.\ 2016, \apj, 823, 19

\bibitem[Dark Energy Survey Collaboration et al.(2016)]{dar16} Dark Energy Survey Collaboration, Abbott, T., Abdalla, F.~B., et al.\ 2016, \mnras, 460, 1270 


\bibitem[Dotter et al.(2008)]{dot08} Dotter, A., Chaboyer, 
B., Jevremovi{\'c}, D., et al.\ 2008, \apjs, 178, 89 
\bibitem[Drlica-Wagner et al.(2015)]{drl15} Drlica-Wagner, A., Bechtol, K., Rykoff, E.~S., et al.\ 2015, \apj, 813, 109 

\bibitem[Fabrizio et al.(2015)]{fab15} Fabrizio, M., Nonino, M., Bono, G., et al.\ 2015, \aap, 580, A18 



\bibitem[Ferguson(1989)]{fer89} Ferguson, H.~C.\ 1989, 
\aj, 98, 367 







\bibitem[Hilker et al.(2003)]{hil03} Hilker, M., Mieske, S., \& Infante, L.\ 2003, \aap, 397, L9 




\bibitem[Jang 
\& Lee(2014)]{jan14} Jang, I.~S., \& Lee, M.~G.\ 2014, \apjl, 795, L6  

\bibitem[Jang \& Lee (2017)]{jan17b} Jang, I.~S., \& Lee, M.~G.\ 2017, \apj, in press (arXiv:161105040) 


\bibitem[Jang et al. (2017)]{jan17} Jang, I.~S., Lee, M.~G. et al.\ 2017, in preparation


\bibitem[Kambas et al.(2000)]{kam00} Kambas, A., Davies, J.~I., Smith, R.~M., Bianchi, S., \& Haynes, J.~A.\ 2000, \aj, 120, 1316


\bibitem[Kim et al.(2015)]{kim15a} Kim, D., Jerjen, H., Mackey, D., Da Costa, G.~S., \& Milone, A.~P.\ 2015, \apjl, 804, L44  
\bibitem[Kim \& Jerjen(2015)]{kim15b} Kim, D., \& Jerjen, H.\ 2015, \apjl, 808, L39 






\bibitem[Koposov et al.(2015)]{kop15}Koposov, S. E., Belokurov, V., Torrealba, G., \& Evans, N. W.
2015, \apj, 805, 130




\bibitem[Laevens et al.(2014)]{lae14} Laevens, B.~P.~M., Martin, N.~F., Sesar, B., et al.\ 2014, \apjl, 786, L3 

\bibitem[Laevens et al.(2015)]{lae15} Laevens, B.~P.~M., Martin, N.~F., Bernard, E.~J., et al.\ 2015, \apj, 813, 44 

\bibitem[Lee et al.(1993)]{lee93} Lee, M.~G., Freedman, W.~L., \& Madore, B.~F.\ 1993, \apj, 417, 553  







\bibitem[Martin et al.(2016)]{mar16} Martin, N.~F., Ibata, R.~A., Lewis, G.~F., et al.\ 2016, arXiv:1610.01158 

\bibitem[McConnachie(2012)]{mcc12} McConnachie, A.~W.\ 2012, 
\aj, 144, 4  





\bibitem[Mieske et al.(2007)]{mie07} Mieske, S., Hilker, M., Infante, L., \& Mendes de Oliveira, C.\ 2007, \aap, 463, 503





\bibitem[Moore et al.(1999)]{moo99} Moore, B., Ghigna, S.,  Governato, F., et al.\ 1999, \apjl, 524, L19  

\bibitem[Mu{\~n}oz et al.(2015)]{mun15} Mu{\~n}oz, R.~P., Eigenthaler, P., Puzia, T.~H., et al.\ 2015, \apjl, 813, L15








\bibitem[Richtler et al.(2012)]{ric12} Richtler, T., Bassino, L.~P., Dirsch, B., \& Kumar, B.\ 2012, \aap, 543, A131



 
 \bibitem[Ricotti et al.(2016)]{ric16} Ricotti, M., Parry, O.~H., \& Gnedin, N.~Y.\ 2016, arXiv:1607.04291 
 




\bibitem[Sand et al.(2012)]{san12} Sand, D.~J., Strader, J., 
Willman, B., et al.\ 2012, \apj, 756, 79  




\bibitem[Schlafly 
\& Finkbeiner(2011)]{sch11} Schlafly, E.~F., \& Finkbeiner, D.~P.\ 2011, \apj, 737, 103 

\bibitem[Schweizer (1980)]{sch80}Schweizer, F. 1980, \apj, 237, 303


\bibitem[S{\'e}rsic(1963)]{ser63} S{\'e}rsic, J.~L.\ 1963, 
Boletin de la Asociacion Argentina de Astronomia La Plata Argentina, 6, 41  



\bibitem[Simon 
\& Geha(2007)]{sim07} Simon, J.~D., \& Geha, M.\ 2007, \apj, 670, 313  



\bibitem[Spitzer(1987)]{spi87} Spitzer, L. 1987, Dynamical evolution of globular clusters, Princeton NJ, Princeton University Press

\bibitem[Stetson(1994)]{ste94} Stetson, P.~B.\ 1994, \pasp, 106, 250  












\bibitem[Willman et al.(2005a)]{wil05a} Willman, B., Blanton, 
M.~R., West, A.~A., et al.\ 2005a, \aj, 129, 2692  


\bibitem[Willman et al.(2005b)]{wil05b} Willman, B., Dalcanton, 
J.~J., Martinez-Delgado, D., et al.\ 2005b, \apjl, 626, L85  






\end{thebibliography}
\end{document}